%% file: samplepaper.tex
\PassOptionsToPackage{table,dvipsnames}{xcolor}
\documentclass[conference]{IEEEtran}
\IEEEoverridecommandlockouts

\usepackage{graphicx}
\usepackage{array}        %
\usepackage{multirow}     %
\usepackage{makecell}     %
\usepackage{url}
\usepackage{float}
\usepackage{booktabs} 
\usepackage{tcolorbox}
\usepackage[table]{xcolor}
\usepackage[dvipsnames]{xcolor}

\newcommand{\up}{\textcolor{teal}{$\uparrow$}}
\newcommand{\dn}{\textcolor{red!70!black}{$\downarrow$}}

\newcommand{\takeaway}[1]{%
  \begin{tcolorbox}[colback=gray!5, colframe=gray!75, boxrule=1.5pt, left=2pt, right=2pt, top=1pt, bottom=1pt]
    \textit{#1}
  \end{tcolorbox}}
  
\begin{document}

\title{Stratifying the Digital Divide:\\Analysis of Socio-Economic Influences on Internet Performance}

\author{
\IEEEauthorblockN{Shivani Kalamadi}
\IEEEauthorblockA{\textit{University of California, Davis}\\
Davis, CA, USA\\
skalamadi@ucdavis.edu}
\and
\IEEEauthorblockN{Aditya Bej}
\IEEEauthorblockA{\textit{University of California, Davis}\\
Davis, CA, USA\\
adityabej1997@gmail.com}
\and
\IEEEauthorblockN{Sachin Kumar Singh}
\IEEEauthorblockA{\textit{University of Utah}\\
Salt Lake City, UT, USA\\
sachinkumar.singh@utah.edu}
\and
\IEEEauthorblockN{Varshika Srinivasavaradhan}
\IEEEauthorblockA{\textit{University of California, Santa Barbara}\\
Santa Barbara, CA, USA\\
varshika@ucsb.edu}
\and
\IEEEauthorblockN{Elizabeth Belding}
\IEEEauthorblockA{\textit{University of California, Santa Barbara}\\
Santa Barbara, CA, USA\\
ebelding@ucsb.edu}
\and
\IEEEauthorblockN{Alexander Gamero-Garrido}
\IEEEauthorblockA{\textit{University of California, Davis}\\
Davis, CA, USA\\
agamerog@ucdavis.edu}
}

\maketitle             

\begin{abstract}
 Despite numerous technological advancements, the digital divide remains a pressing issue affecting millions worldwide. We present a framework for diagnosing internet inequality at the Census Block Group level by pairing approximately 170 million crowdsourced Ookla speed tests (2021--2025) with U.S. Census demographics across six metropolitan regions. 
 After quantifying and correcting for sampling bias, we use Random Forest regression with permutation importance to identify the socio-economic drivers of \textit{download speed}, \textit{upload speed}, and \textit{latency}. 
 Population density dominates all three metrics at the regional level, but this dominance is an artifact of scale: once areas are stratified into density bins, its influence vanishes in medium- and higher-density neighborhoods, revealing that socio-economic conditions are the true differentiators of internet quality in most urban settings. 
 After controlling for density, income and racial composition emerge as the primary drivers, income consistently dictating upload speed and racial composition, proving to be a stronger predictor of download speed than either income or education.
 Our findings demonstrate that internet inequality is locally configured: no single national narrative explains it, and effective policy demands region-specific intervention.

\end{abstract}

\input{introduction}
\input{relatedwork}

\input{datacollection}

\input{method}

\input{results}

\input{discussion}

\input{limitations_conclusion}

\bibliographystyle{IEEEtran}
\bibliography{refs}

\input{appendix}

\end{document}

%% file: introduction.tex
\section{Introduction}

The internet is embedded in nearly every dimension of modern life, from education and healthcare to employment, banking, and government services, all of which increasingly assume reliable broadband as a baseline. The COVID-19 pandemic underscored this dependence: as workplaces, schools, and healthcare moved online almost overnight, residential internet traffic surged by 15--20\% within weeks~\cite{feldmann2020lockdown}, reinforcing that internet access is not a luxury but a fundamental right~\cite{ref_url_humanrights}, and  
whose disparities directly hinder societal participation~\cite{benda2020broadband}.

Yet simply \textit{having} a broadband internet connection is not the same as having a \textit{good} one. In the modern digital economy, the quality of access, measured by \textit{download} and \textit{upload throughput} and \textit{latency}, determines whether a household can sustain a video conference, submit an application, or stream a lecture without disruption. 
This broadband performance varies not only between urban and rural areas~\cite{riddlesden2014broadband} but also across neighborhoods within the same city~\cite{li2023racial, ref_article_racial_analysis}. \textit{Speed} and \textit{reliability} disparities correlate with income~\cite{ref_scienceDirect}, racial composition~\cite{ref_article_racial_analysis}, education levels~\cite{silva2018diversity}, and infrastructure investment patterns, 
with ISPs systematically under investing in low-income and minority communities~\cite{mccall2022socio, skinner2024redlining}.
These disparities persist even within well-connected urban areas~\cite{lee2023analyzing, ref_scienceDirect}, disproportionately affecting certain communities~\cite{ref_article_household_internet_trends}.
The factors behind these disparities vary from place to place, yet most studies examine them at the state or metropolitan level~\cite{ref_article1_paul_performance_inequity, riddlesden2014broadband}, a scale too coarse to capture neighborhood-level variation or to separate the effects of closely correlated socio-economic variables.

Understanding these disparities at a finer granularity is a prerequisite for informed policymaking and targeted intervention~\cite{espin2024bridging,silva2018diversity,tomer2017signs,SAN_ANTONIO_broadband_affordability_deployment}: blanket policies risk misallocating resources while overlooking the communities most in need, and existing federal broadband maps compound the problem by overstating coverage~\cite{gao2018broadband}. 
Achieving finer granularity introduces several methodological challenges.
First, the most widely available large-scale performance data comes from crowdsourced platforms such as Ookla Speedtest~\cite{ref_url_ookla}, which are subject to self-selection and geographic sampling biases that can systematically favor certain populations~\cite{lee2023analyzing}. Second, socio-economic variables such as income, education, race, and age can be highly correlated with one another, making it difficult to isolate the independent contribution of any single factor using standard regression approaches~\cite{ref_education_pays}. Third, population density exerts such a strong influence on broadband availability~\cite{kolko2010new, glass2010empirical, stenberg2009broadband} that it can obscure subtler socio-economic effects unless explicitly controlled for. Without addressing these issues jointly, analyses risk producing findings that are either biased by uneven measurement coverage or confounded by correlated predictors.

We propose a replicable framework that pairs crowdsourced network measurements with Census-level socio-economic data at the Census Block Group (CBG) level, the smallest geographic unit for which the U.S. Census publishes demographic estimates~\cite{ref_url_census}. 
The framework corrects for sampling bias through population-proportional reweighting, employs Random Forest regression~\cite{breiman2001random} with permutation importance to handle multicollinearity among correlated predictors, and stratifies each study area by population density to separate infrastructure-driven effects from socio-economic ones. 
As it relies only on publicly available measurement and demographic data, this framework is not tied to any particular geography and can be applied to other metropolitan, national, or international contexts where there is comparable inputs. 

With this study, we address three research questions:

\begin{itemize}

    \item \textbf{RQ1 (Coverage):} How unevenly are crowdsourced speed tests distributed across regions, neighborhoods, and years and what does this reveal?

\item \textbf{RQ2 (Density):} To what extent does population density confound socio-economic predictors, and how does stratifying by density change what matters?

\item \textbf{RQ3 (Drivers):} After controlling for density, which socio-economic factors shape internet performance, and do they differ across regions?
\end{itemize}

We apply this framework to six U.S. metropolitan regions: the San Francisco Bay Area, Los Angeles, Central California, Dallas, Houston, and San Antonio combining approximately 170 million Ookla speed tests (2021--2025) with Census demographic data across roughly 146,000 CBG-level observations. 
Our analysis reveals that population density dominates internet performance predictions but only in low-density areas, with its influence vanishing entirely in medium-density neighborhoods once stratified and after controlling for density, income and racial composition emerge as the primary drivers, with effects that differ sharply across regions and metrics rather than following a single national pattern.

%% file: relatedwork.tex
\section{Related Work}

\textbf{Broadband Disparities and the Digital Divide:}
A growing body of work documents persistent inequalities in broadband access across the United States. Allen~\cite{ref_article_household_internet_trends} traces national trends from 2013 to 2023, showing that disparities in household internet access remain widespread. Li et al.~\cite{li2023racial} confirm that racial and income gaps in broadband access persist across 905 U.S. cities even within urban areas, while Gallardo and Whitacre~\cite{ref_scienceDirect} find unexpected speed disparities across socioeconomic groups. At the neighborhood level, Rodriguez-Elliott and Vachuska~\cite{ref_article_racial_analysis} show that majority-Black neighborhoods experience lower upload speeds than majority-White neighborhoods, with gaps widening during the COVID-19 pandemic. 

\textbf{Infrastructure Investment and Digital Redlining:}
Disparities in broadband quality are not solely demand-side phenomena. McCall et al.~\cite{mccall2022socio} document how ISPs systematically underinvest in low-income and minority communities, a practice termed digital redlining. Skinner et al.~\cite{skinner2024redlining} link this pattern empirically to historically redlined neighborhoods, where broadband access remains lower despite nominally similar technological availability. Reddick et al.~\cite{SAN_ANTONIO_broadband_affordability_deployment} analyze broadband affordability and access determinants through a community survey in San Antonio, highlighting how local socio-economic context shapes connectivity outcomes. On the supply side, population density is a well-established driver of broadband deployment: denser areas have lower per-household infrastructure costs and stronger broadband supply~\cite{kolko2010new, glass2010empirical, stenberg2009broadband, beede2015broadband}.

\textbf{Crowdsourced Broadband Measurement:}
Crowdsourced speed test platforms, particularly Ookla Speedtest~\cite{ref_url_ookla}, have become a primary data source for large-scale broadband studies. Paul et al.~\cite{ref_article1_paul_performance_inequity} characterize performance inequity across U.S. Ookla users, while Lee et al.~\cite{lee2023analyzing} analyze temporal progression of internet quality with bias correction. MacMillan et al.~\cite{macmillan2023comparative} compare Ookla with M-Lab's NDT7, and Alford-Teaster et al.~\cite{alford2024broadband} examine spatiotemporal variation in download/upload metrics for telehealth accessibility. However, crowdsourced data suffers from well-documented sampling biases, favoring urban, higher-income users that can distort aggregated metrics~\cite{lee2023analyzing, ref_article1_paul_performance_inequity}. Prior work addresses this through population-proportional reweighting, an approach we adopt for all six of our study regions.

\textbf{Our Work:}
Prior studies have documented broadband disparities across demographic dimensions but typically operate at coarse geographic scales, rely on pairwise correlations, or examine a single predictor in isolation. Our work operates at the Census Block Group level across six metropolitan regions, using Random Forest regression with permutation importance and null model validation to handle multicollinearity and isolate independent feature contributions. 
By stratifying each region by population density, we uncover that density's dominance is an artifact of scale --- it vanishes in medium- and higher-density neighborhoods, where socio-economic conditions become the true differentiators, a pattern that prior aggregate-level analyses have been unable to detect.

%% file: datacollection.tex
\section{Data Collection}
\label{sec:data}
We focus on six U.S. metropolitan regions, chosen to capture region-wide broadband trends across diverse demographic and infrastructural contexts: San Francisco Bay Area, Los Angeles, Central California, Dallas, Houston, and San Antonio. These represent the largest metropolitan regions within the two most populous states of the U.S. (Appendix~\ref{sec:counties} lists the counties comprising each region). To ensure our analysis includes both urban cores and lower-density suburban areas, we considered entire counties encompassing these regions, not just individual municipalities. 

\subsection{Ookla Measurement Data}
Internet experience is tightly coupled with key network performance metrics such as throughput (download/upload) and latency. Many online speed-test platforms allow users to measure these metrics directly from their devices. These speed tests typically select a nearby measurement server based on the client’s location and round-trip time (RTT), and then use multiple parallel TCP connections to saturate the bottleneck link and estimate throughput. In this paper, we use data from Ookla Speedtest~\cite{ref_url_ookla}, which are derived from user-initiated, crowdsourced tests collected when an end user chooses to run a measurement. Crowdsourced measurements from Ookla Speedtest have been widely used in prior work~\cite{ref_article1_paul_performance_inequity, lee2023analyzing,gallardo2024unexpected,
macmillan2023comparative,alford2024broadband,
sanchez2023understanding,
bauer2010understanding,
canadi2012revisiting}. 
This dataset is particularly well-suited for our research since they capture real-world internet performance as experienced by end users across diverse socio-economic contexts, at a scale and geographic granularity that no active probing infrastructure could feasibly replicate.

We utilize Ookla Fixed measurement data from 2021 to 2025, encompassing approximately 170 million raw speed tests across our six studied regions. We aggregate this data to the CBG level to align with Census demographic boundaries, yielding approximately 146K data points that we use for our analysis. Each data point in our dataset thus represents multiple aggregated speed tests, with the number of contributing tests varying across regions and years. For instance, the average tests per CBG in the Bay Area are nearly double those in Central California.

To maintain statistical reliability, we include only those CBGs with at least 20 recorded tests in our analysis. This threshold is empirically motivated: CBGs with fewer than 20 tests exhibit a coefficient of variation in download speed exceeding 0.87, indicating high measurement instability, while CBGs with 20 or more tests show a coefficient of variation below 0.66, representing the largest single-step reduction in measurement instability across all test count brackets that we selected. CBGs with a tests-to-devices ratio exceeding 25 were also excluded since values above this threshold, as identified from a natural break in the global ratio distribution at the 99.99th percentile, indicate automated or institutional testing infrastructure rather than residential broadband usage. The average Spearman correlation between the total number of tests and the total number of devices is $> 0.92$, indicating a strong positive relationship. This suggests that regions with more tests typically have proportionally more unique devices, reducing the risk of overrepresentation by a small set of users within a CBG and lending credibility to the aggregated performance metrics.

\subsection{Census Data}
We pair the Ookla measurements with yearly demographic data from the U.S. Census Bureau's American Community Survey (ACS)~\cite{ref_url_census}, available at the CBG level. This data includes variables such as population density, median household income, gender distribution, education attainment levels, age brackets, and racial composition. Since the latest ACS data is not yet available for the 2025 dataset, we use 2024 demographic estimates. 

\begin{figure}[H]
\includegraphics[width=\columnwidth]{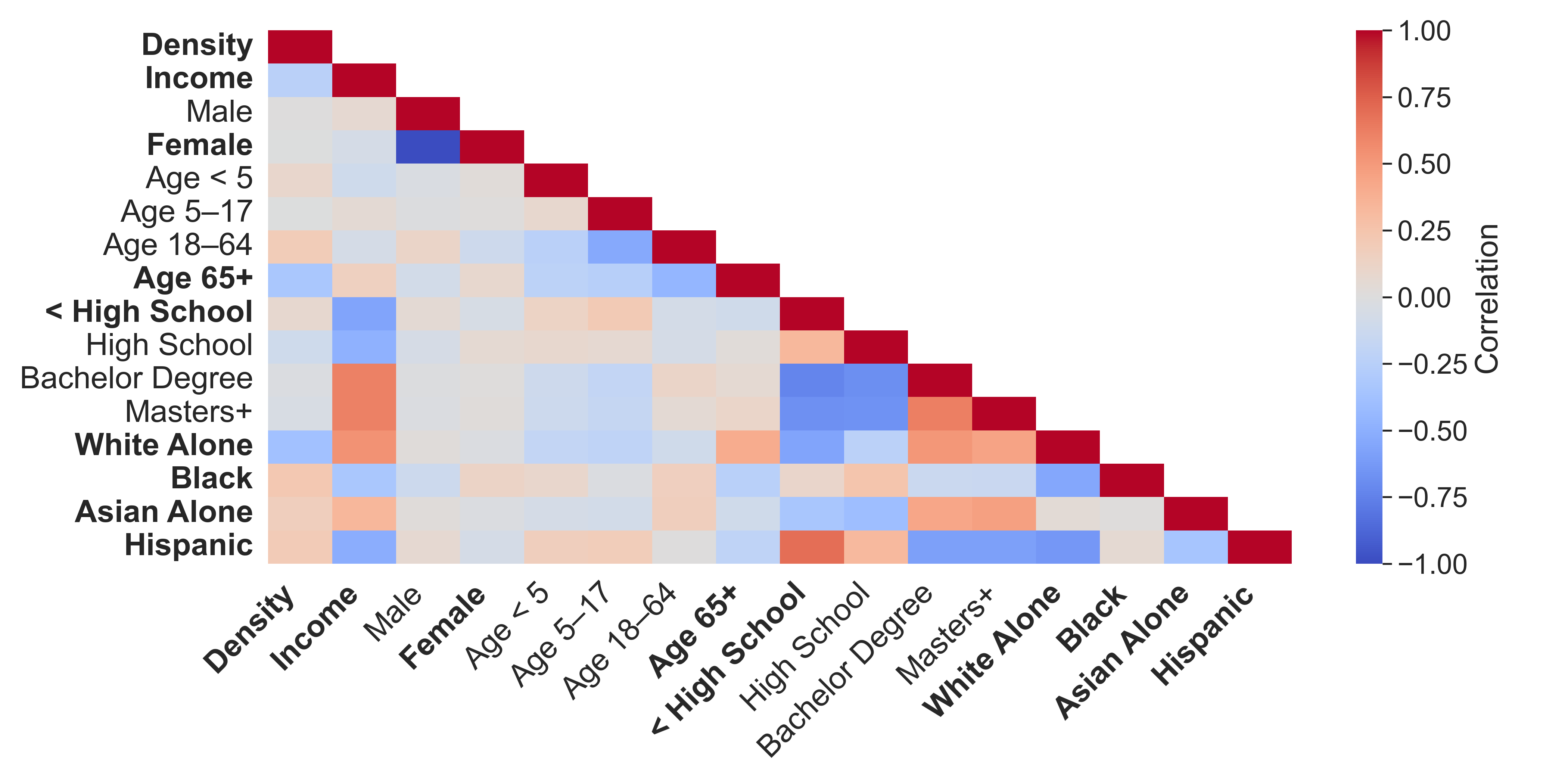}
\caption{Correlation between socio-economic variables in Dallas.} 
\label{heatmap}
\end{figure}
Socio-economic variables are often interdependent: income frequently co-varies with educational attainment~\cite{ref_education_pays} and race, as illustrated by the correlation matrix for Dallas in Figure~\ref{heatmap}. 
These inter-dependencies make it challenging to isolate the effect of any single variable on broadband performance, as apparent relationships may instead be driven by correlated features. Moreover, the regional variability in these correlations underscores the need to analyze at an intermediate geographic scale, which is broad enough to capture structural trends, yet granular enough to reveal local disparities.

\subsubsection{Feature Selection}
Apart from \textit{Population Density}, \textit{Income} and Gender: \textit{Percentage Female}, we limit our model inputs to a concise, representative subset of features to mitigate the effects of the correlations described earlier. Within each demographic category, we select a single variable, since the constituent percentages sum to the total population and are therefore linearly dependent. For age, we select the \textit{percentage of population aged 65 or above}, which has been shown to face inequities in internet access ~\cite{ref_pew2017,ref_friemel2016}. Among education attributes, we retain \textit{less than a high school diploma}, which consistently appeared as a strong predictor in preliminary model runs. For race, which cannot be reduced to a single proxy, we include the most populous groups: \textit{White alone}, \textit{Black or African American alone}, \textit{Asian alone}, and \textit{Hispanic/Latino}.

 This yields nine CBG-level features that we use throughout the paper to contextualize broadband performance and as inputs to our predictive models of \textit{download speed}, \textit{upload speed}, and \textit{latency}.

%% file: method.tex
\section{Methodology}

Our pipeline proceeds in three stages. \textit{First}, we detect and correct for sampling bias in the crowdsourced measurements, a well-established step in studies using Ookla data~\cite{ref_article1_paul_performance_inequity, lee2023analyzing}, to ensure population-representative performance estimates \textbf{(RQ1)}. \textit{Second}, we fit a Random Forest regression model with permutation importance and null model validation to identify which socio-economic features reliably predict internet performance, and apply filtering criteria to retain only robust associations \textbf{(RQ2, RQ3)}. \textit{Third}, we stratify each region by population density to separate infrastructure-driven effects from socio-economic ones, revealing the predictors that emerge once density is controlled for \textbf{(RQ2, RQ3)}.

\subsection{Sampling Bias Detection and Correction}
\label{sec:bias}
Crowdsourced speed tests like Ookla’s can be affected by self-selection (\emph{e.g.}, users testing during a connection problem), uneven test distribution driven by device usage, user behavior, and regional broadband access. Accordingly, we treat the dataset as observational and use methods~\cite{ref_article1_paul_performance_inequity, lee2023analyzing} to mitigate the impact of these sampling biases.

\begin{table}[H]
\centering
\caption{Region-wise sampling bias: percent difference between
expected tests based on population share and observed Ookla
Speedtests, 2025.}
\label{tab:sampling_bias}
\setlength{\tabcolsep}{8pt}
\begin{tabular}{lrrr}
\toprule
\textbf{Region} & \textbf{Population} & \textbf{Observed Tests} & \textbf{Difference} \\
\midrule
Bay Area            & 7,618,903  & 6,789,068  & 26.78\% \\
Los Angeles         & 18,481,956 & 12,782,687 & 54.70\% \\
Central California  & 3,578,287  &   988,265  & 31.62\% \\
Dallas              & 7,985,590  & 4,279,023  & 18.78\% \\
Houston             & 7,442,788  & 3,684,791  & 15.15\% \\
San Antonio         & 2,662,733  & 1,021,774  & 17.72\% \\
\bottomrule
\end{tabular}
\end{table}
While the tests-to-devices ratio described in Section~\ref{sec:data} validates the distribution of measurements \textit{within} individual CBGs, it does not address whether tests are distributed proportionally \textit{across} CBGs relative to population.
Following prior work~\cite{ref_article1_paul_performance_inequity, lee2023analyzing}, 
we use a chi-squared test to compare observed versus expected test counts based on population share, which confirmed sampling bias in all regions of our dataset. For example, Los Angeles shows the largest discrepancy at 54.70\%, with only 12.8 million observed tests against an expected share based on its 18.5 million population, while Houston and San Antonio show smaller but still significant gaps of 15.15\% and 17.72\% respectively (Table~\ref{tab:sampling_bias}).

To correct for this, we apply Cumulative Distribution Function (CDF) to adjust the influence of each test by reweighing test counts in proportion to the population size of the corresponding CBG. 
Each CBG's contribution is adjusted by upweighting under-sampled areas and downweighting over-sampled areas relative to their population share. This produces population-representative estimates of internet performance for each region. The effect of this correction on test count distributions is presented in Section~\ref{sec:rq1}.

\subsection{Feature Importance Model using Random Forest}

We use a tree-based regression model to determine the most important socio-economic features in each region. Random Forests are ensemble learning methods that are known to handle multicollinearity well~\cite{breiman2001random,
strobl2008conditional}, capture non-linear relationships, and offer interpretable rankings of feature contributions against dependent variables (download speed, upload speed, and latency).

A Random Forest is a collection of many decision trees, each trained on a slightly different sample of the data. Each tree recursively partitions the feature space by selecting the feature and split threshold that minimizes the weighted mean squared error (MSE) across the resulting child nodes. The feature importance score for feature is then computed as the normalized total reduction in MSE attributable to splits across all trees, which makes the model more stable and better able to generalize than a single decision tree. This also allows the model to provide feature importance scores, which indicate how much each socio-economic variable contributes to improving prediction accuracy across the forest.
Scores are normalized to sum to 1 across all features. A higher score indicates that the feature contributes more to reducing prediction error across the forest.

Hyperparameters were selected via grid search with 5-fold cross-validation, evaluating combinations of number of trees ($T \in \{100, 200, 500\}$), 
maximum depth ($\{10, 20, 30\}$), and minimum samples per split ($\{2, 5, 10\}$). The configuration $T = 200$, maximum depth $= 20$, and minimum samples per split $= 5$ minimized mean absolute error across held-out folds and was applied uniformly across all region-year-metric-bin combinations.

\subsection{Permutation Importance Model}

Although the Random Forest model provides an initial ranking of feature importance, its predictive errors remain relatively high, suggesting that the relationship between socio-economic variables and internet performance is noisy and cannot be captured reliably by standard feature importance scores alone. To make the analysis more robust, we therefore complement this approach with Permutation Importance.

Permutation Importance works by randomly shuffling the values of a single feature (e.g., income) and measuring the resulting decrease in model accuracy. Features that, when shuffled, cause a larger drop in accuracy are considered more important, as they would be better able to determine the value of our dependent variable. In other words, we judge a feature’s importance by how much model performance worsens when that feature’s information is disrupted. This makes the results more stable and easier to interpret, because it measures how much each feature actually helps the model make accurate predictions.

\subsection{Null Model (Baselines)}

We also used a Null Model as a validation mechanism to benchmark the performance of our feature-based predictions against random baseline estimates. Rather than shuffling individual feature values as in permutation importance, the null model randomly shuffles the 
dependent variable: download speed, upload speed, or latency, while leaving all features intact. This severs any real relationship between the socio-economic predictors and the network metric, so the resulting feature importance scores represent what the model would produce under pure chance.

Null model extends the Permutation approach since Permutation importance shuffles one socio-economic feature at a time, meaning the scores it produces still reflect the influence of all other features in the model. A feature that appears important under permutation may simply be because of another correlated predictor. Null model overcomes this by breaking all feature-outcome relationships simultaneously, producing a true random baseline against which each permutation score can be compared.

\subsection{Filtering criteria}
\label{sec:filtering}

A feature may appear important in a given bin simply because the model itself has weak predictive power, or because the permutation score does not meaningfully exceed what would be expected by chance. 
To ensure reliability, we apply three criteria before interpreting any result, all of which must be satisfied simultaneously:

\begin{enumerate}
    \item \textbf{Model fit:} The bin-level model must achieve a coefficient of determination $R^2 \geq 0.10$. Bins where the model explains less than 10\% of the variance are excluded, as feature rankings from such models are unreliable.
    \item \textbf{Effect size:} The permutation importance of a feature must exceed 0.10, ensuring that only features with a non-trivial contribution to model accuracy are retained.
    \item \textbf{Above baseline:} The permutation importance must exceed the corresponding null model importance by at least 0.10, confirming that the feature's contribution is meaningfully above the random baseline.
\end{enumerate}

\subsection{Interpretation}
For each region-year-metric-bin combination, we identify a set of features that pass all three filtering criteria. Two quantities guide how we interpret each surviving feature.

\begin{enumerate}
\item \textbf{Magnitude}: A permutation importance score reflects the proportional drop in model accuracy when a feature's values are randomly shuffled. A score of 0.10, for instance, indicates that disrupting that feature reduces predictive  accuracy by approximately 10\%. Higher scores therefore indicate stronger, more reliable predictors of the network metric in the given region.

\item \textbf{Direction}: Permutation importance quantifies how much a feature matters but not whether the relationship is positive or negative. We determine direction using the Spearman correlation coefficient, which captures rank-based monotonic relationships without assuming linearity, with Pearson correlation used as secondary confirmation.

\end{enumerate}

\subsection{Density-Based Stratification}
We begin our modeling at the region level (unstratified), where population density consistently emerges as the dominant predictor across all six regions and all five years. This finding is consistent with prior work showing that population density is a fundamental driver of broadband availability and deployment, since denser areas typically have lower per-household infrastructure costs and stronger broadband supply~\cite{kolko2010new, glass2010empirical, beede2015broadband, stenberg2009broadband}.

Because density's dominance can mask the influence of other socio-economic features, we use it as a stratification variable rather than a competing predictor. 
We divide the CBGs of each region into five equal-sized bins based on population density. Within each bin, density variation is controlled, allowing the model to surface whichever socio-economic features best explain the remaining variation in internet performance. This stratification captures how the identity and relative influence of predictors shift across low- to high-density neighborhoods, a pattern that would be invisible at the aggregate regional level.

%% file: results.tex
\section{Results}
\label{sec:results}

\subsection{\textbf{RQ1: Measurement Coverage and Bias}}
\label{sec:rq1}

\subsubsection{\textbf{Test Count Trends}}
\begin{figure}[t]
\centering
\includegraphics[width=0.8\columnwidth]{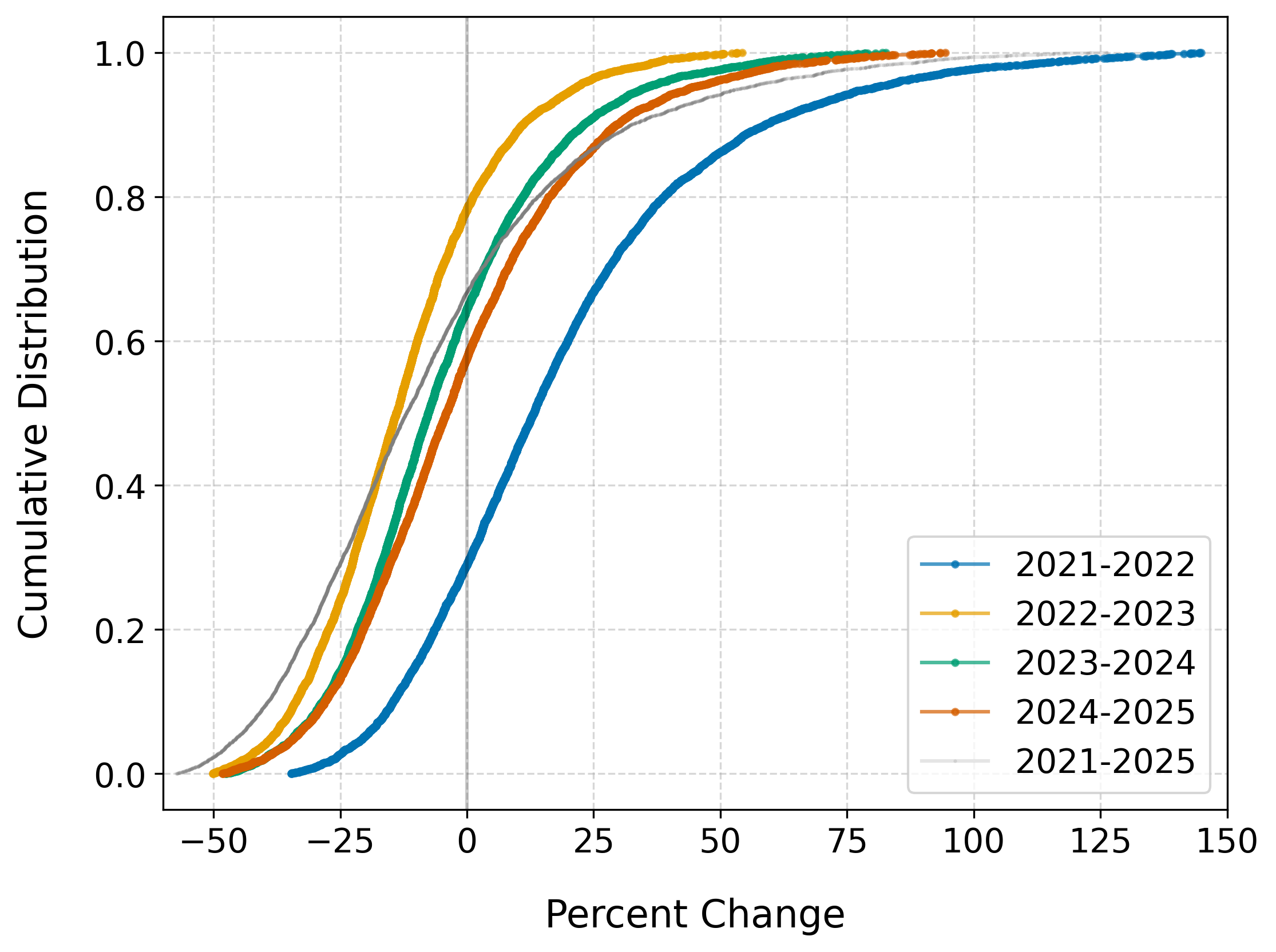}
\caption{Percent Change in the number of tests in Bay Area by year.} 
\label{bayarea_tests_cdf}
\end{figure}

The distribution of year-over-year percent change in test counts across CBGs reveals a clear temporal arc in testing behavior. In the Bay Area (Figure~\ref{bayarea_tests_cdf}), the 2021--2022 period shows the largest surge in testing activity, likely driven by post-pandemic remote work. Subsequent years show progressive stabilization, with the 2024--2025 distribution nearly flat. This surge-then-stabilization pattern is consistent across all six regions, revealing structural unevenness in measurement coverage that reflects engagement disparities across communities.

\subsubsection{\textbf{Effect of Bias Correction}}
As described in Section~\ref{sec:bias}, a chi-squared test confirmed statistically significant sampling bias in all six regions. 
These differences indicate that certain communities are systematically over- or under-represented in the crowdsourced data relative to their population share.
\begin{figure*}[!t]
\includegraphics[width=\textwidth]{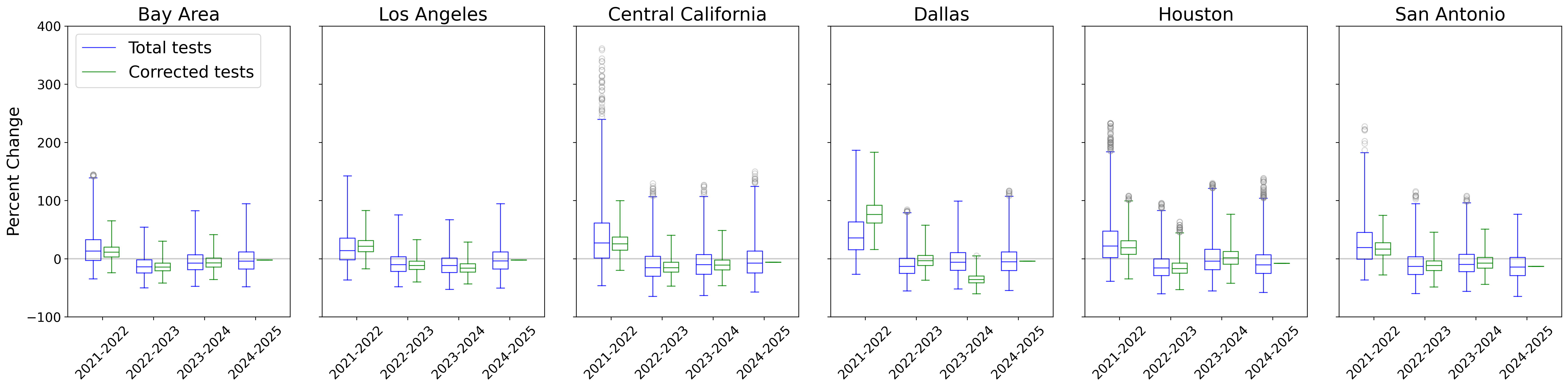}
\caption{Variation in number of tests before and after sampling bias correction. Whiskers extend to $3 \times \mathrm{IQR}$; top and bottom 1\% of values trimmed to account for the long tail distribution} 
\label{boxplots_testcount_percentange}
\end{figure*}
Figure~\ref{boxplots_testcount_percentange} shows the year-over-year percent change in test counts before and after bias correction across all six regions. 
Across all regions, the corrected test distributions show notably tighter interquartile ranges and shorter whiskers than their uncorrected counterparts, indicating that reweighting reduces the variability introduced by uneven testing behavior.

The magnitude of correction varies by region. Central California exhibits the most extreme uncorrected spread, with outliers exceeding 300\% year-over-year change in 2021--2022, reflecting highly uneven test coverage across its CBGs. Houston shows a similarly wide uncorrected distribution in the same period. In both cases, the corrected distributions collapse to a narrow band near zero, demonstrating that the population-proportional reweighting effectively absorbs these imbalances. By contrast, regions like the Bay Area and Los Angeles show smaller uncorrected spreads to begin with, suggesting more organically balanced test coverage, though the correction still produces a measurable tightening.

These results confirm that crowdsourced speed test data cannot be taken \textit{as-is} for neighborhood-level analysis. The systematic differences between observed and expected test counts, if left uncorrected, would bias any downstream inference toward the demographics and neighborhoods that test most frequently. All subsequent analyses in this paper use the corrected, population-reweighted data.

\takeaway{\textbf{Takeaway RQ1:} Crowdsourced speed test data exhibits significant and uneven sampling bias across all six regions, with discrepancies between observed and expected test counts ranging from 15\% to 55\%. Testing behavior follows a post-pandemic surge-then-stabilization arc, and without population-proportional reweighting, downstream analyses would systematically favor neighborhoods that test most frequently.}

\subsection{\textbf{RQ2: Population Density Dominance and Stratification}}
\label{sec:rq2}
We run our analysis on six U.S. regions against our network measurement metrics: Download Speed, Upload Speed, and Latency. This leads to 6 regions × 5 years × 3 metrics × 9 features each = 810 results. 
After applying our filtering criteria (Section~\ref{sec:filtering}), 91 unique validated trends remain. We present these results organized by research question.
\subsubsection{\textbf{Unstratified Analysis}}

\begin{figure}[H]
\includegraphics[width=\columnwidth]{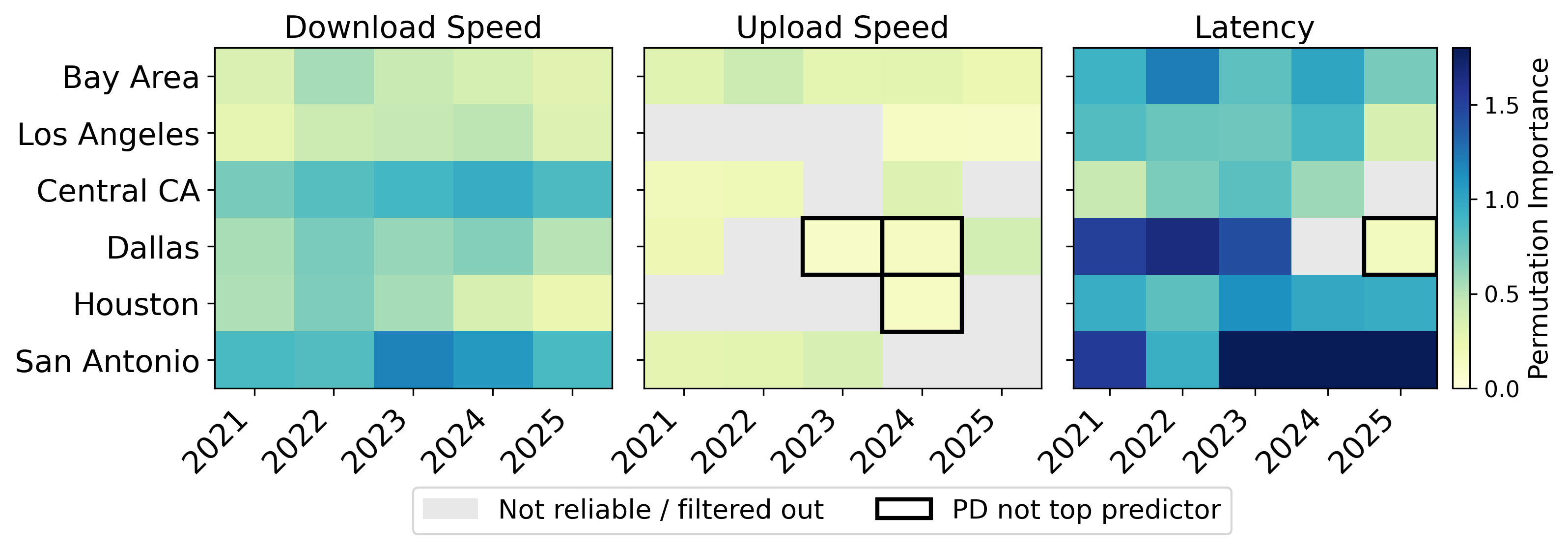}
\caption{Population density permutation importance across all regions (unstratified). Gray cells did not meet filtering criteria and black borders mark combinations where population density (PD) was not the top predictor.}
\label{pd_heatmap_all_bin}
\end{figure}

Without density stratification, population density emerges as the dominant predictor across all six regions and all five years as shown in Figure \ref{pd_heatmap_all_bin}. 
For \textit{download speed}, it ranks as the top validated feature in all 30 region-year combinations, with Spearman correlation confirming a consistently positive relationship i.e., higher density predicts faster download speeds everywhere. Signal strength varies considerably across regions: Central California and San Antonio show the strongest associations (permutation importance up to 0.957 and 1.182 respectively), while the Bay Area shows weaker but consistent effects. 

For \textit{latency}, population density is not only dominant but exhibits the strongest permutation importance scores of any metric, as reflected by the consistently dark cells in Figure~\ref{pd_heatmap_all_bin}. It leads in 27 of 28 reliable combinations, with the sole exception being Dallas 2025 (where Hispanic/Latino share emerges as the top predictor). This likely reflects the tight coupling between population density and proximity to network infrastructure, which directly affects \textit{latency}.

\textit{Upload speed} presents a different picture. Population density leads where results are reliable, but a large proportion of region-year combinations are filtered out entirely (gray cells in Figure~\ref{pd_heatmap_all_bin}), indicating that population density alone often fails to produce a result meeting our three thresholds for reliability. 
Where reliable models do exist, population density is not always the top predictor (the black-bordered cells in Figure~\ref{pd_heatmap_all_bin}), notably surpassed by median household income in Dallas (2023--2024) and Houston (2024), foreshadowing the socio-economic patterns examined in Section~\ref{sec:rq3}.
Los Angeles (2024–2025) presents a distinct case where population density remains the top upload predictor but with a negative Spearman coefficient confirmed by Pearson, suggesting that in the densest urban areas of LA, density may reflect infrastructure saturation or network congestion rather than expanded service availability.

\subsubsection{\textbf{Stratified Analysis}}

Since population density dominates the unstratified analysis, for subsequent analyses, we implement population density–based bucketing to validate our findings across varying density levels within each region. Table~\ref{tab:reliability} summarizes the stratified results.

\begin{table}[H]
\centering
\caption{Filtered results and population density dominance 
across density bins.}
\label{tab:reliability}
\begin{tabular}{lrrr}
\toprule
\textbf{Density Bin} & \textbf{Validated Findings} & \textbf{PD Dominant} & \textbf{PD \%} \\
\midrule
Low Density         & 80 & 56/63 & 89\% \\
Medium-Low Density  & 16 &  1/10 & 10\% \\
Medium Density      &  13 &  0/8  &  0\% \\
Medium-High Density & 15 &  0/12 &  0\% \\
High Density        & 28 &  2/22 &  9\% \\
\bottomrule
\end{tabular}
\end{table}

The transition from low to medium-low density reveals a sharp decline rather than a gradual decline: population density accounts for 89\% of reliable combinations in the lowest density bin (56 of 63) but drops to just 10\% in the medium-low bin (1 of 10) and reaches 0\% in both the medium and medium-high bins. This abrupt falloff suggests a threshold effect: once a neighborhood reaches a moderate level of density, infrastructure availability becomes sufficiently uniform that density no longer differentiates internet performance, and socio-economic characteristics take over as the primary drivers.

High-density areas partially recover with 28 validated findings, but population density accounts for only 2 of 22 (9\%) of these. The features driving the remaining 20 associations are socio-economic, indicating that in the densest urban neighborhoods, variation in internet quality is shaped by who lives there rather than by infrastructure access alone.

The number of validated feature associations (i.e., reliable feature-metric relationships passing all three reliability filter criteria)  also shifts substantially across bins: low-density areas yield 80 findings, reflecting strong and consistent signals, while medium density bins produce far fewer reliable results (13-16), suggesting that at intermediate density levels, fewer feature dominates consistently. High-density areas partially recover with 28 validated findings, but these are driven mostly by socio-economic features rather than infrastructure access. Importantly, each of these 152 validated findings across the stratified bins represents a trend unique to a specific region, year, metric, and density context.

\takeaway{\textbf{Takeaway RQ2:} 
Population density overshadows other factors, but only in low-density neighborhoods. Once stratified into equal-sized density bins, its dominance disappears in medium and higher density areas, where socio-economic features emerge as the primary drivers of internet performance.
}

\subsection{\textbf{RQ3: Socio-Economic Drivers After Controlling for Density}}
\label{sec:rq3}

In medium to high density bins, where population density no longer dominates, socio-economic and demographic features emerge as reliable predictors of network performance. In this analysis, we restrict to the single top-ranked socio-economic feature per region-year-metric combination to avoid conflation across overlapping signals. Table~\ref{tab:socio_predictors} summarizes the most frequent validated features across the Medium-low bin through the highest bin, ranked by occurrence count. Direction is determined by the Spearman correlation coefficient between each feature and the corresponding network metric.

\subsubsection{\textbf{Aggregate Patterns}}
Median household income is the most frequent predictor (9 upload speed associations, all positive). Racial composition accounts for 12 of 15 download speed findings, but with non-uniform directions: White and Asian shares associate with \textit{lower} speeds while Hispanic/Latino share associates with \textit{higher} speeds. For latency, Black or African American share appears in 4 combinations, all indicating worse performance. The regional decomposition below examines whether these patterns are geographically uniform or concentrated.

\subsubsection{\textbf{Regional Decomposition}} 
\begin{figure}[t]
\includegraphics[width=\columnwidth]{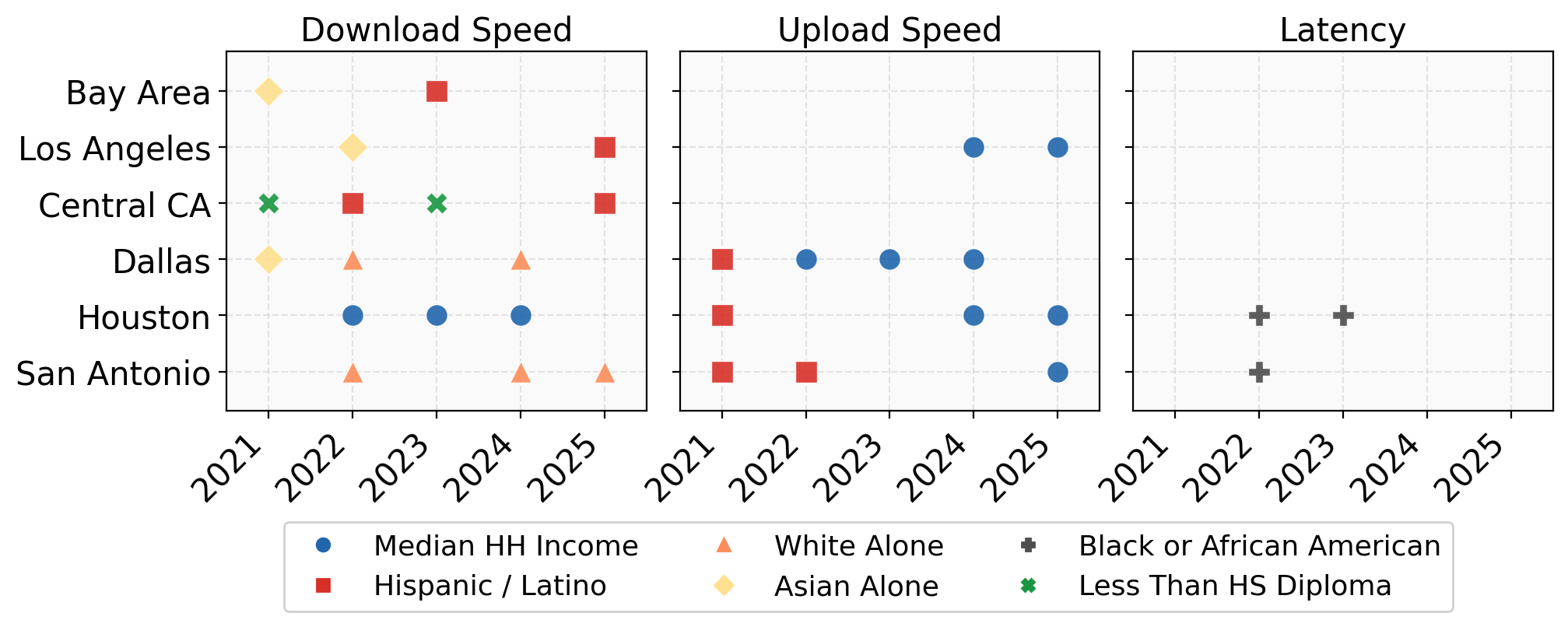}
\caption{Top-ranked socio-economic predictor for each validated region-year-metric combination after density-based stratification.}
\label{socio_economic_features_dotgrid}
\end{figure}

A critical question is whether these patterns reflect uniform trends across all six regions or are concentrated in specific geographic contexts.
Figure~\ref{socio_economic_features_dotgrid} visualizes where each validated socio-economic finding originates. In Table~\ref{tab:socio_predictors}, each count represents unique region-year-bin combinations whereas the graph shows unique region-year combinations, so counts may differ.

\paragraph{\textbf{Median Household Income}} Income emerges as the dominant upload speed predictor almost exclusively in Texas: 6 of the 8 validated income-upload associations originate from Dallas and Houston, with Dallas showing a consistent signal across 2022-2024 and Houston appearing in 2024-2025. This temporal consistency spanning multiple consecutive years within the same region strengthens the case that this is a stable structural relationship rather than incidental. For download speed, income surfaces only in Houston 2022-2024, again in consecutive years, suggesting that in Houston's denser neighborhoods, household income constrains both upload 
and download performance. Income exhibits no significant association with latency in any of the cities analyzed in this study.

\begin{table}[t]
\centering
\caption{Most frequent socio-economic predictors in medium and high 
density bins, ranked by occurrence.}
\label{tab:socio_predictors}
\begin{tabular}{llrr}
\toprule
\textbf{Feature} & \textbf{Metric} & \textbf{Dir.} & \textbf{Count} \\
\midrule
\rowcolor{gray!15} Median Household Income   & Upload Speed   & \up & 9 \\
\rowcolor{gray!15} Hispanic Latino           & Upload Speed   & \up & 6 \\
White Alone               & Download Speed & \dn & 5 \\
Hispanic Latino           & Download Speed & \up & 4 \\
Less Than High School     & Download Speed & \dn & 4 \\
\rowcolor{gray!30} Black or African American & Latency        & \up & 4 \\
Asian Alone               & Download Speed & \dn & 3 \\
Median Household Income   & Download Speed & \up & 3 \\
\bottomrule
\end{tabular}
\end{table}

\paragraph{\textbf{Racial Composition}} Racial trends are geographically split in ways 
that reveal distinct regional dynamics rather than a uniform pattern.

\textit{White Alone} predicts lower download speeds in 5 validated combinations, and every one of these originates from Texas. This association does not surface in any California region, suggesting that the relationship between White residential concentration and download speed may be mediated by Texas-specific infrastructure deployment or neighborhood investment patterns rather than race alone.

\textit{Hispanic/Latino} share predicts network metrics in 8 validated combinations, and 4 of these are concentrated in California, with no equivalent signal in Texas. Despite Texas having substantial Hispanic populations, the effect does not extend to download speeds. However, early Texas regions (2021 and 2022) show a relationship with Hispanic population and upload speeds with speeds increasing with the demographics. 

\textit{Asian Alone} predicts lower download speeds in 3 validated combinations, appearing in Bay Area (2021), Los Angeles (2022) and Dallas (2021). Critically, this signal does not extend beyond these two early years and it does not resurface, suggesting it may reflect infrastructure conditions that have since changed rather than a persistent structural disparity.

\textit{Black or African American} share predicts worse latency in 3 validated combinations, concentrated in Houston (2022-2023) and San Antonio (2022). This signal is geographically limited to Texas and temporally clustered around the same years.

\paragraph{\textbf{Educational Attainment}} The less-than-high-school-diploma feature surfaces as the top download speed predictor in only one region across the entire study period: Central California, in 2021 and 2023. The fact that this signal appears in two non-consecutive years within the same region suggests a recurring rather than incidental relationship, but its absence from all other regions indicates that the education-connectivity link, while documented in prior literature, may be observable at the CBG level only in areas with particularly sharp educational stratification such as Central California's agricultural communities.

\takeaway{\textbf{Takeaway RQ3:} No single socio-economic factor explains internet performance uniformly. Income drives upload speed in Texas, racial composition drives download speed across regions, and the direction of these effects varies by geography. Internet inequality is locally configured.
}

%% file: discussion.tex
\section{Discussion}

\paragraph*{\textbf{Race as a Regionally Specific Predictor}} Our results indicate that the racial composition of a region can significantly influence internet access, but the effects are nuanced and vary across regions. 
For instance, in San Antonio 2025, the mean White population share across CBGs is 26.3\%. Consistent with our model findings, CBGs below this average exhibit a higher mean download speed of 407 Mbps, whereas CBGs above the average show a lower mean download speed of 375.45 Mbps. Another example involves Hispanic/Latino population share in Los Angeles 2025, where the mean Hispanic share across CBGs is 47.8\%. Here, CBGs above this average show a higher mean download speed of 376.38 Mbps compared to 350.88 Mbps for CBGs below the average. 
These examples illustrate that the relationship between racial composition and download speed varies in both direction and magnitude across regions, consistent with the finding that racial predictors are geographically specific rather than nationally generalizable~\cite{SAN_ANTONIO_broadband_affordability_deployment}.

\paragraph*{\textbf{Predictors Do Not Always Apply Locally}}
Despite prior work finding age significant at national scales~\cite{ref_pew2017, ref_friemel2016}, neither age nor gender survives our filtering criteria. These null findings demonstrate that predictors identified as significant at national scales do not necessarily translate to neighborhood-level analysis, reinforcing that the drivers of internet performance are not necessarily the same as those operating at the neighborhood level, and that scale-appropriate variable selection is essential for accurate diagnosis.

\paragraph*{\textbf{The Case for CBG-Level Resolution}} In LA counties alone, the CBG-level download speeds range from 42 to 584 Mbps in 2025, demonstrating that county-level averages would entirely obscure the disparities we identify. 
Our findings operate at the neighborhood scale and can only be detected through granular CBG-level analysis, reinforcing that meaningful broadband equity assessment requires fine-grained geographic resolution and that policies targeting county or state-level aggregates risk misallocating resources by averaging over communities with fundamentally different connectivity realities.

\paragraph*{\textbf{Toward Region-Specific Policy.}}
Because the dominant predictor changes by region, metric, and density, our findings argue against one-size-fits-all broadband policy. Upload disparities in Texas call for affordability interventions such as tiered subsidies. 
Download disparities tied to racial composition call for infrastructure audits to identify neighborhood-level deployment gaps. Our framework serves as a diagnostic tool for identifying \textit{which} intervention is appropriate \textit{where}.

\paragraph*{\textbf{Scope of Top-Feature Reporting.}}
Our results report only the single most important feature per region-year-metric-bin combination. A more comprehensive understanding of internet access disparities requires examining secondary and tertiary features, which may reveal meaningful interaction effects. Future work employing feature interaction analysis could decompose these layered effects more finely.

%% file: limitations_conclusion.tex
\section{Limitations}
\label{sec:limitations}

Our findings should be interpreted in light of the following limitations inherent to the data and study design.

\paragraph{Crowdsourced speed-test data} Our reweighting adjusts for how many tests come from each CBG, but not for who within a CBG is running them, meaning any systematic link between demographic characteristics and testing propensity - such as higher-income residents testing more frequently - propagates into our corrected sample.

\paragraph{Geographic} Findings are specific to six U.S. metropolitan regions and may not generalize to rural areas or other metros, though the framework itself  is transferable.

\paragraph{Associative rather than causal inference.} 
Our models identify associations, not causes. Highly ranked features may proxy unmeasured confounders such as ISP market concentration or fiber deployment history. Identifying these would require an experimental approach that is beyond the scope of this study. Our findings characterize \textit{where} and \textit{which} 
demographic dimensions co-vary with internet performance, not \textit{why}.

\section{Conclusion}

Our study provides a novel, replicable framework that uncovers evidence of internet inequality associated with socio-economic features that go beyond the traditionally recognized uniform influences of population density or income. 
We show that population density's influence is confined to low-density areas and vanishes entirely once neighborhoods reach moderate density, where income, racial composition, and education become the operative drivers. 
Because these drivers differ by region, metric, and even direction, no single broadband policy can address all communities and neighborhood equally. 
Built on publicly available crowdsourced measurement and Census data, our framework can be applied to any geography where comparable inputs exist, offering policymakers a diagnostic tool for identifying not just where internet inequality concentrates but which socio-economic dimensions drive it locally.

%% file: appendix.tex
\section{Appendix}

\subsection{Counties included per region}
\label{sec:counties}
\begin{table}[h]
\centering
\scriptsize
\setlength{\tabcolsep}{3pt}
\begin{tabular}{ll}
\toprule
\multicolumn{2}{c}{\textbf{California}} \\
\midrule
\textit{Bay Area:} & Alameda, Contra Costa, Marin, Napa, San Mateo, Santa Clara, \\
& San Francisco, Solano, Sonoma \\[4pt]
\textit{Los Angeles:} & Los Angeles, Orange, Riverside, San Bernardino, Ventura \\[4pt]
\textit{Central CA:} & Fresno, Kern, Kings, Madera, Mariposa, Merced, Stanislaus, Tulare \\
\midrule
\multicolumn{2}{c}{\textbf{Texas}} \\
\midrule
\textit{Dallas:} & Collin, Dallas, Denton, Ellis, Hunt, Johnson, Kaufman, Parker, \\
& Rockwall, Tarrant, Wise \\[4pt]
\textit{Houston:} & Austin, Brazoria, Chambers, Fort Bend, Galveston, Harris, Liberty, \\
& Montgomery, San Jacinto, Waller \\[4pt]
\textit{San Antonio:} & Atascosa, Bandera, Bexar, Comal, Guadalupe, Kendall, Medina, Wilson \\
\bottomrule
\end{tabular}
\end{table}